\documentstyle[twocolumn,aps]{revtex}

\begin{document}
\draft
\title{Nonlinear microscopic relaxation of uniform magnetization precession }
\author{Vladimir L. Safonov and H. Neal Bertram}
\address{Center for Magnetic Recording Research, \\
University of California - San Diego, \\
9500 Gilman Drive, La Jolla, CA 92093-0401}
\date{\today}
\maketitle

\begin{abstract}
Dynamic relaxation for nonlinear magnetization excitation is analyzed. For
direct processes, such as magnon-electron scattering and two-magnon
scattering, the relaxation rate is determined from the linear case simply by
utilizing the magnetization oscillation frequency for nonlinear excitation.
For an indirect process, such as slow-relaxing impurities, the analysis
gives an additional relaxation term proportional to the excitation level. In
all cases the effective magnetization damping is increased compared to
Landau-Lifshitz-Gilbert damping.
\end{abstract}


\section{Introduction}

The study of damped magnetization dynamics of fine ferromagnetic particles
and thin films is important for the development of nanomagnetic devices and
high-density magnetic recording. A conventional theoretical tool to study
magnetization relaxation is based on the phenomenological Landau-Lifshitz
equation \cite{LL} or its modification with Gilbert damping \cite{gilbert}.
These equations conserve the absolute value of magnetization ($\left| {\bf M}%
\right| ={\rm const}${\rm )} in a single domain region due to a strong
exchange interaction approximation.\ They are relatively simple and
therefore have been utilized for various calculations and micromagnetic
simulations. However both Landau-Lifshitz and Gilbert equations were
introduced (a) for small magnetization motions and (b) for the case of high
magnetic symmetry (axial symmetry) with an isotropic damping fitting
parameter $\alpha $ (``damping constant''). Nevertheless both ``a'' and
``b'' conditions are usually violated.

Recently a theoretical approach \cite{safonov}, \cite{safbertthermreversal}, 
\cite{safbertbook}, \cite{wangbertsaf}, \cite{wangbertsaf1} has been
developed to correct the limitations of the Landau-Lifshitz-Gilbert (LLG)
theory. The main idea was to represent the magnetization dynamics as the
motion of a damped nonlinear oscillator with the random force of thermal
fluctuations. The oscillator model is a convenient tool to establish a
``bridge'' between the microscopic physics, where the rotational oscillator
variables $a^{\ast }$ and $a$ naturally describe spin excitations (as
creation and annihilation operators) and the macroscopic magnetization
dynamics. In particular, it has been rigorously shown by including specific
coupling of a magnetic system to a variety of loss mechanisms, that for
small oscillations near equilibrium the macroscopic damping term reflects
the anisotropy of the system \cite{BertSafJin}, \cite{safbertrelax}. Only
for complete uniaxial symmetry do the LLG results apply.

Nonlinearity in the relaxation process appears with the increase of
magnetization deviation from equilibrium. Depending on the situation, the
nonlinear magnetization damping either increases or decreases the total
relaxation rate. For example, in our simulations of magnetization reversal
(large magnetization motion) by spin-wave dynamics in a quasi-single-domain
grain \cite{simulation} we have demonstrated that an effective damping
parameter $\alpha $ can be about hundred times greater than that in the
linear regime. For a single-domain grain a phenomenological extension of the
oscillator approach to nonlinear excitations has been made \cite{safonov}.
In this case for large amplitudes the effective damping \ is much greater
than that of LLG.

The aim of this paper is to give a microscopic analysis of nonlinear
magnetization damping. We consider both direct and indirect loss mechanisms
and show explicitly how nonlinearity is included in the relaxation rate. The
approach is to first transform the magnetization dynamics without damping to
rotational oscillator coordinates. This transformation yields equations
where the oscillator frequency depends on the degree of nonlinearity. The
analysis parallels the approach for low-level excitations.

\section{The nonlinear oscillator model}

Here we review the magnetization dynamics in terms of a nonlinear
oscillator. For simplicity we shall consider a single-domain grain in the
case when the external magnetic field ${\bf H}_{0}$ is parallel to the
uniaxial anisotropy axis ($z$). To describe the magnetization dynamics ${\bf %
M(}t{\bf )}$, we introduce the classical spin ${\bf S}={\bf M}V/\hbar \gamma 
$, where $V$ is the grain volume, $\hbar $ is Planck's constant\ and $\gamma 
$\ is the gyromagnetic ratio. Using the Holstein-Primakoff transformation:

\begin{eqnarray}
S^{+} &=&a\,\sqrt{2S-N},\quad S^{z}=S-a^{\ast }a,  \label{hoprim} \\
S^{-} &=&a^{\ast }\,\sqrt{2S-N},\quad S^{\pm }=S^{x}\pm iS^{y},  \nonumber
\end{eqnarray}
where $N=a^{\ast }a=|a|^{2}$ ($N\leq 2S$), we can represent the
magnetization dynamics in terms of the oscillator variables:

\begin{equation}
da/dt=-i\widetilde{\omega }(N)a.  \label{nondynamics}
\end{equation}
The oscillator (\ref{nondynamics}) is characterized by the magnetization
oscillation frequency for nonlinear excitation: 
\begin{equation}
\widetilde{\omega }(N)=\omega _{0}[1-N/S(1+H_{0}/H_{K})],  \label{nonlinfreq}
\end{equation}
which reflects the change of effective magnetic field with increasing
magnetization deviation from equilibrium. $H_{K}$ is the anisotropy field, $%
\omega _{0}=\gamma (H_{0}+H_{K})$\ is the ferromagnetic resonance frequency.

Previously it was assumed that the corresponding stochastic differential
equation is of the following form \cite{safonov}, \cite{safbertbook}:

\begin{equation}
\left( d/dt+\eta (N)\right) a=-i\widetilde{\omega }(N)a+f(t),
\label{dynamic}
\end{equation}
where $f(t)$ describes a random uncorrelated force and $\eta (N)$ is the
nonlinear damping. The formulas (\ref{hoprim}) and (\ref{nonlinfreq}) are
valid within one potential well up to the top of the energy barrier $0\leq
N<N_{top}$, where $\widetilde{\omega }(N)=0$ ($N_{top}=S(1+H_{0}/H_{K})$).
Utilizing the fluctuation-dissipation theorem for (\ref{dynamic}), the
nonlinear damping is given by \cite{safonov}:

\begin{equation}
\eta (N)=\alpha \tilde{\omega}(N),\quad \alpha =\eta (0)/\omega _{0}.
\label{Qfactor}
\end{equation}

\section{Nonlinear relaxation mechanisms}

In this section we show how nonlinear relaxation arises from microscopic
mechanisms. The linear relaxation rate $\eta (0)$ is well understood and
determined by solving the coupled magnetic - thermal bath equations. The
main mathematical idea is that the principal complex amplitude dynamics in
the small damping approximation ($\eta (N)\ll \widetilde{\omega }(N)$) is
given (from either (\ref{nondynamics}) and (\ref{dynamic})) by 
\begin{equation}
a(t)\propto \exp [-i\widetilde{\omega }(N)t].  \label{nonamplit}
\end{equation}
For all cases where the relaxation is of a direct form of coupling (see, 
\cite{safbertrelax}), the microscopic analysis will yield the same
relaxation expression for a given mechanism except that $\omega _{0}$ is
replaced by $\widetilde{\omega }(N)$.

\subsection{Magnon-electron scattering}

Let us consider the magnon-electron scattering process in a ferromagnetic
metal \cite{safbertrelax}, \cite{KamberskyPatton}. A magnon with wave vector 
${\bf k}=0$ and energy $\hbar \omega _{0}$ and a conduction electron with
wave vector ${\bf k}\neq 0$ and energy $\hbar \omega _{e{\bf k}}$ are
transformed into a conduction electron with wave vector ${\bf k}^{\prime }$
and energy $\hbar \omega _{e{\bf k}^{\prime }}=\hbar \omega _{e{\bf k}%
}+\hbar \omega _{0}$. This confluence process occurs in the presence of
defects to violate momentum conservation. The linear relaxation rate for
this process can be written as:

\begin{equation}
\eta _{m-e}=c_{def}\pi \sum_{{\bf k},{\bf k}^{\prime }}\left| \frac{f_{{\bf %
kk}^{\prime }}}{N}\right| ^{2}(\overline{n}_{{\bf k}}-\overline{n}_{{\bf k}%
^{\prime }})\delta (\omega _{e{\bf k}^{\prime }}-\omega _{e{\bf k}}-\omega
_{0}),  \label{KP14}
\end{equation}
where $c_{def}$ is the defect concentration, $f_{{\bf kk}^{\prime }}/N$ is
the scattering amplitude and $\overline{n}_{{\bf k}}$ is the Fermi
occupation number. Neglecting $\omega _{0}$ dependence in the scattering
amplitude and taking into account that $\omega _{0}\ll \omega _{e{\bf k}}$
and $\overline{n}_{{\bf k}}-\overline{n}_{{\bf k}^{\prime }}\simeq -\omega
_{0}\partial \overline{n}_{{\bf k}}/\partial \omega _{e{\bf k}}$, from (\ref
{KP14}) we obtain:

\begin{equation}
\eta _{m-e}=c_{def}\omega _{0}\pi \sum_{{\bf k}}\left( -\frac{\partial 
\overline{n}_{{\bf k}}}{\partial \omega _{e{\bf k}}}\right) \sum_{{\bf k}%
^{\prime }}\left| \frac{f_{{\bf kk}^{\prime }}}{N}\right| ^{2}\delta (\omega
_{e{\bf k}^{\prime }}-\omega _{e{\bf k}})  \label{kampa}
\end{equation}
with a linear dependence $\eta _{m-e}\propto \omega _{0}$. Replacing $\omega
_{0}\rightarrow \widetilde{\omega }(N)$, we have:

\begin{equation}
\eta _{m-e}(N)/\widetilde{\omega }(N)=\eta _{m-e}/\omega _{0}=\alpha _{m-e}
\label{meprocess}
\end{equation}
in agreement with the phenomenological relation (\ref{Qfactor}).

\subsection{Two-magnon scattering}

The linear regime of two-magnon scattering on defects has been considered by
many authors (e.g., \cite{sparks}, \cite{gurevich}, \cite{twomagnon}, \cite
{AriasMills}). The linear relaxation rate for this process is given by:

\begin{equation}
\eta _{2m}=\pi \sum_{{\bf k}}|G_{{\bf k}}|^{2}\delta (\omega _{{\bf k}%
}-\omega _{0}),  \label{twomagnon}
\end{equation}
where $G_{{\bf k}}$ describes the scattering amplitude and $\omega _{{\bf k}%
} $ is the spin-wave frequency with ${\bf k\neq }0$. The abovementioned
simplification gives:

\begin{equation}
\eta _{2m}(N)=\pi \sum_{{\bf k}}|G_{{\bf k}}|^{2}\delta \left[ \omega _{{\bf %
k}}(N)-\widetilde{\omega }(N)\right] .  \label{2mnonlinear}
\end{equation}
Here $\omega _{{\bf k}}(N)$ is the spin-wave frequency taking account of
nonlinear excitation. This example does not give, in general, a linear
relation between $\eta _{2m}(N)$ and $\widetilde{\omega }(N)$.

\subsection{Slow relaxation}

The energy loss in this case occurs via an intermediate damped dynamic
system, the `slow relaxing' impurities (e.g., \cite{safbertrelax}, \cite
{gurevich}, \cite{mikhailov}). The magnetization motion modulates the
impurity splittings and varies the thermal equilibrium populations of the
energy levels. Let us consider two-level impurities with energy:

\begin{equation}
{\cal H}_{imp,j}=\hbar \sum_{j}\left[ \Omega _{0,j}+\delta \Omega _{j}(t)%
\right] n_{j},  \label{hamimpu}
\end{equation}
where $\Omega _{0,j}$ is the splitting frequency, $n_{j}$ is the upper lever
population and $j$ is the impurity index. The impurity level modulation is
defined as:

\begin{equation}
\delta \Omega _{j}(t)=\Phi _{j}a(t)+\Phi _{j}^{\ast }a^{\ast }(t).
\label{implevelmod}
\end{equation}

The nonlinear dynamic equation for the complex amplitude is:

\begin{eqnarray}
da/dt &=&-i\widetilde{\omega }(N)a-i\partial ({\cal H}_{imp,j}/\hbar
)/da^{\ast }  \label{dyneqnmi} \\
&=&-i\widetilde{\omega }(N)a-i\sum_{j}\Phi _{j}^{\ast }\delta n_{j}(t). 
\nonumber
\end{eqnarray}
The kinetics of the impurity population is defined by the following equation:

\begin{equation}
dn_{j}/dt=-\Gamma _{\parallel ,j}(\Omega _{j})[n_{j}-n_{T}(\Omega _{j})].
\label{impurity}
\end{equation}
Here $\Gamma _{\parallel ,j}(\Omega _{j})$ is the impurity relaxation rate
and $n_{T}(\Omega _{j})=[\exp (\hbar \Omega _{j}/k_{B}T)+1]^{-1}$ is the
equilibrium population at frequency $\Omega _{j}(t)=\Omega _{0,j}+\delta
\Omega _{j}(t)$. We can we solve Eq.(\ref{impurity}) and obtain:

\begin{eqnarray}
\delta n_{j}(t) &=&\int\limits_{-\infty }^{t}\exp [\phi _{\parallel
,j}(t_{1})-\phi _{\parallel ,j}(t)]~\Gamma _{\parallel ,j}[\Omega
_{j}(t_{1})]  \label{general} \\
&&\times \{n_{T}[\Omega _{j}(t_{1})]-n_{T}(\Omega _{0,j})\}dt_{1},  \nonumber
\end{eqnarray}
where $\delta n_{j}(t)=n_{j}(t)-n_{T}(\Omega _{0,j})$ and $\phi _{\parallel
,j}(t)=\int\limits_{-\infty }^{t}\Gamma _{\parallel ,j}[\Omega
_{j}(t^{\prime })]dt^{\prime }$.

In the case of small modulation $\hbar |\delta \Omega _{j}(t)|/k_{{\rm B}%
}T\ll 1$ we can write the following expansion: $n_{T}[\Omega _{0,j}+\delta
\Omega _{j}(t)]=n_{T}(\Omega _{0,j})+[\partial n_{T}(\Omega _{0,j})/\partial
\Omega _{0,j}]\delta \Omega _{j}(t)+...$ Neglecting for simplicity the\ $%
\Omega _{j}(t)$ dependence of $\Gamma _{\parallel ,j}$, we can rewrite Eq.(%
\ref{general}) in terms of $a(t)$ and $a^{\ast }(t)$ and substitute these
terms into (\ref{dyneqnmi}). Thus, we can obtain the nonlinear relaxation
rate in the form:

\begin{equation}
\eta _{sr}(N)\simeq \eta _{sr}+\eta _{sr}^{(1)}N,  \label{srnonlinear}
\end{equation}
where

\begin{equation}
\eta _{sr}\simeq \sum\limits_{j}\left| \Phi _{j}\right| ^{2}\left( -\frac{%
\partial n_{T}(\Omega _{0,j})}{\partial \Omega _{0,j}}\right) \frac{\omega
_{0}\Gamma _{\parallel ,j}}{\Gamma _{\parallel ,j}^{2}+\omega _{0}^{2}}
\label{SlowRelaxRate}
\end{equation}
is a linear relaxation rate and

\begin{equation}
\eta _{sr}^{(1)}\simeq \sum\limits_{j}\frac{\left| \Phi _{j}\right| ^{4}}{2}%
\left( -\frac{\partial ^{3}n_{T}(\Omega _{0,j})}{\partial \Omega _{0,j}^{3}}%
\right) \frac{\omega _{0}\Gamma _{\parallel ,j}}{\Gamma _{\parallel
,j}^{2}+\omega _{0}^{2}}  \label{sr-non}
\end{equation}
is the coefficient of a nonlinear damping. We can see that, even though the
relaxation rate depends on the level of excitation, in general, $\eta
_{sr}(N)/\widetilde{\omega }(N)\neq \eta _{sr}/\omega _{0}$.

\section{Discussion}

We have considered nonlinear relaxation for both direct and indirect
coupling to a thermal bath. For direct coupling it was shown that the linear
relaxation rate can be simply converted to the nonlinear rate by replacing
the linear frequency $\omega _{0}$ by the nonlinear frequency $\widetilde{%
\omega }(N)$. In one case of direct coupling, magnon-electron scattering,
the nonlinear relaxation rate was shown to be directly proportional to the
nonlinear frequency, as in the initial phenomenological approach (\ref
{Qfactor}). For the case of two-magnon scattering the relaxation rate, in
general, is not proportional to the nonlinear frequency. A specific case of
indirect interaction, `slow-relaxing' impurities, was analyzed in detail.
The resulting relaxation rate was the sum of the linear term plus a
nonlinear term proportional to the level of excitation. An absence of linear
relation between the nonlinear damping $\eta (N)$ and magnetization
oscillation frequency for nonlinear excitation $\widetilde{\omega }(N)$
indicates a colored thermal noise in the system.

In Ref.\cite{safonov} it was shown that the nonlinear oscillator damping is
greater than the conventional damping in the Landau-Lifshitz equation. This
result is a general conclusion for all the relaxation mechanisms considered
here. Corresponding to (\ref{dynamic}) and (\ref{Qfactor}) the magnetization
dynamic equation can be written as

\begin{equation}
d{\bf M}/dt=-\gamma {\bf M}\times {\bf H}_{{\rm eff}}+(\widetilde{\alpha }%
/M_{s}){\bf M}\times d{\bf M}/dt,  \label{principal}
\end{equation}
where, in general, 
\begin{equation}
\widetilde{\alpha }=\frac{\eta (N)}{\widetilde{\omega }(N)(1-N/2S)},\quad 
\frac{N}{S}=\frac{M_{s}-M^{z}}{M_{s}}=1-\cos \theta .  \label{alphasq}
\end{equation}
Here $\theta $ is the deviation angle. In the vicinity of equilibrium ($N=0$%
, $M^{z}=M_{s}$, $\theta =0$) $\widetilde{\alpha }=\alpha =\eta (0)/\omega
_{0}$, where $\alpha $ is the LLG parameter. Away from equilibrium, the
damping parameter increases $\widetilde{\alpha }$ with increasing
nonlinearity, becoming infinite as $N\rightarrow 2S$ ($M^{z}\longrightarrow
-M_{s}$, or $\theta \longrightarrow \pi $). An application of this approach
to magnetization reversal is discussed in Ref.\cite{WangBertram}.

\section{Acknowledgment}

This work was partly supported by matching funds from the Center for
Magnetic Recording Research at the University of California - San Diego and
CMRR incorporated sponsor accounts.

\end{document}